\documentclass[apex,twocolumn]{jjap3}
\usepackage{txfonts,graphicx,bm,color}
\usepackage{amssymb,amsmath}

\title{Half-integer Shapiro-steps in superconducting qubit with a 
$\pi$-Josephson junction}

\author{Michiyasu Mori$^{1}$ and Sadamichi Maekawa$^{1,2,3}$}
\inst{$^{1}$Advanced Science Research Center, Japan Atomic Energy Agency, 
	Tokai, Ibaraki 319-1195, Japan\\
$^{2}$RIKEN Center for Emergent Matter Science, Wako 351-0198, Japan\\
$^{3}$Kavli Institute for Theoretical Sciences, University of Chinese 
Academy of Sciences, Beijing 100049, China}
\abst{
	A superconducting quantum interference device (SQUID) comprising 0- and 
	$\pi$-Josephson junctions (JJs), called $\pi$-SQUID, is studied by the 
	resistively 
	shunted junction model. The $\pi$-SQUID shows half-integer Shapiro-steps (SS) 
	under 
	microwave irradiation at the voltage $V$ = $(\hbar/2e)\Omega (n/2)$, with 
	angular frequency $\Omega$ and half-integer $n$/2 in addition to integer $n$. 
	We show that the $\pi$-SQUID 
	can be a $\pi$-qubit with spontaneous loop currents by which the half-integer 
	SS are induced. Making the 0- and $\pi$-JJs equivalent is a key for the 
	half-integer 
	SS 
	and realizing the $\pi$-qubit.
}

\date{\today}

\begin{document}
\maketitle

A superconducting quantum interference device (SQUID) shows steps in the 
current-voltage ($I$$-$$V$) curve under microwave irradiation. 
This feature called Shapiro-steps (SS) appears at the voltage 
$V$ = $n$($\hbar$/2$e$)$\Omega$ with an angular frequency of microwave 
$\Omega$, 
integer $n$, the Planck constant $h=2\pi\hbar$ and the elementary charge 
$e$\cite{shapiro,Baro,Tink}. 
Because the frequency of the microwave and the fundamental constants are 
precisely 
determined, the voltage can be defined in the order of 10$^{-9}$ 
accuracy\cite{hamilton,kholmann}. 
The magnetic flux through the SQUID is quantized using $\Phi_0$ $\equiv$ 
$h/(2e)$. 
When a magnetic field with the magnitude of $\Phi_0/2$, which is equivalent to 
the $\pi$-shift, is externally applied, the SS appear at 
half-integer multiples of $V$=$(n/2)$($h$/2$e$)$\Omega$ in addition to the integer 
multiples\cite{vanneste}.

In a Josephson junction (JJ) separated by a ferromagnet, i.e., superconductor/ferromagnet/superconductor (SFS) junction, 
the current-phase relation is shifted by $\pi$ compared with the conventional JJ, and it is called $\pi$-JJ\cite{bulaevskii,buzdin82,buzdin05,bergeret}. 
In an SFS junction, SS can be induced by ferromagnetic 
resonance\cite{hikino08}, and it is sensitive to a domain wall motion in the 
ferromagnet\cite{hikino12,mori}. 
Half-integer SS are also observed in the SFS junction with a 
thickness modulation\cite{frolov,weides,pfeiffer,goldobin} and in the grain boundary of high-$T_C$ 
cuprates\cite{early}. 
This phenomenon originates from the inductive coupling between a $0$-JJ and a $\pi$-JJ\cite{luca,Romeo04,Romeo05}.  
Such a SQUID, including the $\pi$-JJ, is called $\pi$-SQUID, which is realized using a high-$T_c$ cuprate as well\cite{shulz,lindstrom}. 

A solid-state qubit is a key to realize quantum computers\cite{nakamura99,mooij,nakamura01,nakamura03}, and 
decoupling from the environment is essential\cite{ioffe99}. The $\pi$-SQUID is a potential candidate for a $\lq\lq$quiet" qubit not using an external field\cite{blatter,yamashita05,yamashita06}; it is called $\pi$-qubit. 
There are some proposals, such as two JJs with a 0-JJ and a metallic 
$\pi$-JJ\cite{yamashita05}, and three JJs with two 0-JJs and one 
$\pi$-JJ\cite{yamashita06}. 
The Rabi-oscillation in a $\pi$-qubit comprising a SFS junction is observed\cite{feofanov}. 
Because of the $\pi$-JJ, the $\pi$-qubit can form a quantum two-level system without an external magnetic field. 

In this letter, we show that the $\pi$-qubit using the $\pi$-SQUID is 
characterized by the half-integer SS, which are optimized by 
making the 0- and $\pi$-JJs equivalent in terms of critical current and 
resistivity. 
Making two JJs in $\pi$-SQUID equivalent yields a $\pi$-qubit.

\begin{figure}[htbp]
	\centering
	\includegraphics[width=0.45\textwidth]{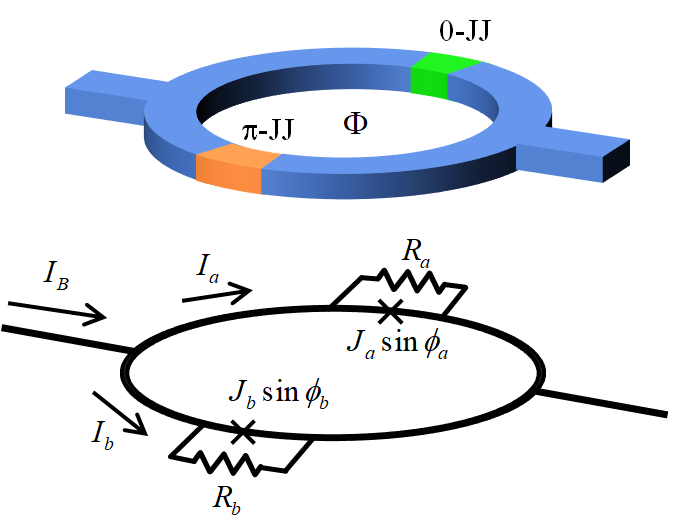}
	\caption{Schematic of $\pi$-SQUID (upper) and the RSJ (lower) models.}
	\label{circuit}
\end{figure}

A schematic of the $\pi$-SQUID comprising 0- and $\pi$-JJs is shown in Fig. \ref{circuit}. 
It can be studied using the resistively shunted junction (RSJ) model composed of two parallel circuits of JJs\cite{vanneste}.
\begin{align}
	I_j&=\frac{V_j}{R_j}+J_j\sin\phi_j,\label{rsj}\\
	\frac{d\phi_j}{dt}&=\frac{2\pi}{\Phi_0} V_j,\label{jjrel}\\                
	\phi_a-\phi_b&=\frac{2\pi}{\Phi_0}\left(\Phi-\Phi_\pi\right),\label{4}\\
	\Phi&=\Phi_{ex}+\frac{L}{2}\left(I_b-I_a\right),\label{flux}\\
	I_a+I_b&=I+I_{ac}\cos\Omega t \equiv I_B,\label{6} 
\end{align}
where $j$ = $a, b$. 
Each junction comprises a resistor $R_j$ and a Josephson current 
$J_j\sin\phi_j$, with current $I_j$, voltage $V_j$, and a phase difference 
$\phi_j$ at the $j$-JJ. 
In Eq. (\ref{flux}), the total flux through the SQUID is denoted by $\Phi$ = $n 
\;\Phi_0$, which is the sum of
an externally applied flux $\Phi_{ex}$ and the flux induced by a screening current. 
The additional flux $\Phi_{\pi}$ = $\Phi_0/2$ is due to the $\pi$-JJ in $\pi$-SQUID and $\Phi_{\pi}$ = 0 in the conventional one.  
Notably, we consider the overdamped case, i.e., with negligible capacitance.
Considering the Josephson relation, Eq. (\ref{jjrel}), 
the equation of motion for phase differences $\phi_a$ and $\phi_b$ are given by,
\begin{align}
\frac{d\phi_a}{d\tau} 
+\sin\phi_a
+\frac{1}{\beta}\left(\phi_a-\phi_b\right)
	&=
	\frac{1}{2}\left[i_B
	-\frac{4\pi}{\beta}\frac{\Phi_\pi-\Phi_{ex}}{\Phi_0}
	\right],\label{eqm9}\\
\frac{d\phi_b}{d\tau}
+\sin\phi_b
-\frac{1}{\alpha \beta }\left(\phi_a-\phi_b\right)
	&=
	\frac{1}{2\alpha}\left[i_B
	+\frac{4\pi}{\beta}\frac{\Phi_\pi-\Phi_{ex}}{\Phi_0}
	\right].\label{eqm10}
\end{align}
where $R_a J_a = R_b J_b \equiv R J$, $\omega_0$  $\equiv$ $2\pi R J/\Phi_0$, $\tau$ $\equiv$ $\omega_0 t$, $\beta$  $\equiv$ $2\pi L J_a/\Phi_0$, $i$ $\equiv$ $I/J_a$, $i_{ac}$ $\equiv$ $I_{ac}/J_a$, and $i_B$ $\equiv$ $I_B/J_a$. 
As we will discuss later, one of the key parameters is $\alpha$ $\equiv$ $J_b/J_a$, which indicates the asymmetry of two JJs.

\begin{figure}[htbp]
	\centering
	\includegraphics[width=0.5\textwidth]{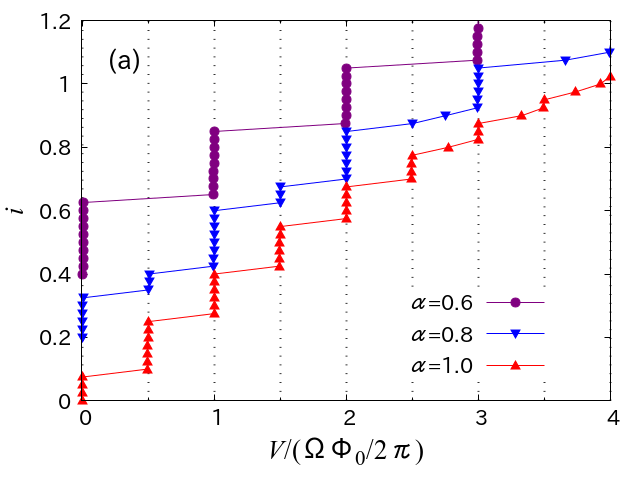}
		\includegraphics[width=0.5\textwidth]{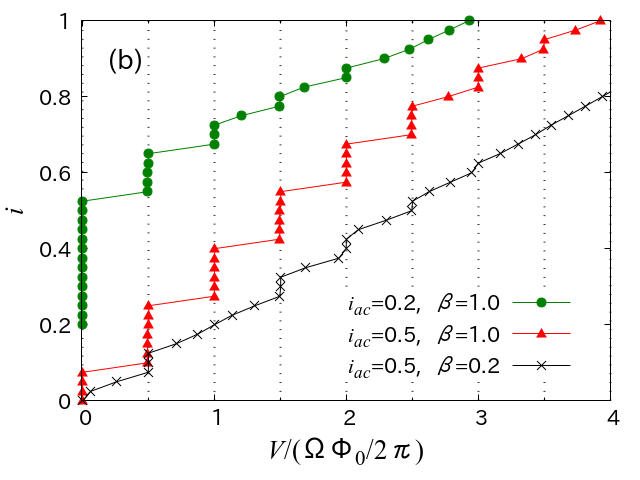}
	\caption{(a) $I$$-$$V$ curves with $\Omega/\omega_0=0.1$, $\beta=1.0$, and $i_{ac}=0.5$, for 
		$\alpha=1.0$ (red upper triangles),  
		$\alpha=0.8$ (blue lower triangles), and 
		$\alpha=0.6$ (purple circles).
		For clarity, the latter two curves are vertically shifted by $i_{ac}$ = 0.2 and 0.4, respectively. 
		(b) $I$$-$$V$ curves with $\Omega/\omega_0=0.1$ and $\alpha=1.0$, for $i_{ac}$ = 0.2 and $\beta$ = 1.0 (green circles), 
		$i_{ac}$ = 0.2 and $\beta$ = 1.0 (red upper triangles), 
		$i_{ac}$ = 0.5 , and $\beta$ = 0.2 (black crosses). The first curve (green 
		circles) is s
		vertically shifted by $i_{ac}$ = 0.2 for clarity.}
	\label{iv}
\end{figure}

By numerically solving Eqs. (\ref{eqm9}) and (\ref{eqm10}) for $\Phi_{ex}=0$, $I$$-$$V$ curves with step structures are obtained as shown in Figs. \ref{iv} (a) and (b). 
In Fig. \ref{iv} (a) for $\beta=1.0$ and $i_{ac}=0.5$, $I$$-$$V$ curves are plotted for 
$\alpha$ = 1.0 (red upper triangles),  
$\alpha$ = 0.8 (blue lower triangles), and 
$\alpha$ = 0.6 (purple circles). 
For clarity, the latter two curves are vertically shifted by $i_{ac}$ = 0.2 and 0.4, respectively.
The half-integer SS are suppressed by decreasing $\alpha$, 
which is controlled by changing the ratio of junction areas, $W_a$ and $W_b$.  
Making both Josephson coupling and resistance of two JJs equivalent is a key 
to observe the half-integer SS. 
In Fig. \ref{iv} (b) for $\alpha$ = 1, $I$$-$$V$ curves are plotted for 
$i_{ac}$ = 0.2 and $\beta$ = 1.0 (green circles), 
$i_{ac}$ = 0.5 and $\beta$ = 1.0 (red upper triangles), and  
$i_{ac}$ = 0.5 and $\beta$ = 0.2 (black crosses). 
The height of half-integer SS is enhanced by increasing $i_{ac}$, 
whereas it is suppressed by decreasing $\beta$. 
When $\beta$ is small by decreasing the SQUID loop, the half-integer Shapiro 
steps can be observed by increasing the $i_{ac}$. 
$\beta$ is estimated as $\beta\sim$ 1 for 2.5$\times$2.5 $\mu$m$^2$ loop and $J_a\sim$70 $\mu$A\cite{vanneste},  
meaning $L\sim$ 4.7 pH. It satisfies the criteria to overcome the thermal noise, i.e., $L$ must be less than about 20 nH at 4.2 K\cite{Tink}. 

The half-integer SS can be understood using the following approximation. In the first order 
of $\beta$,
$i_B$ is given by,
\begin{align}
	i_B
	&\sim 4\frac{\alpha}{1+\alpha} \left[ \cos\left(\pi\frac{\Phi_{ex}-\Phi_\pi}{\Phi_0}\right) \sin\phi 
	+ \frac{\beta}{2}\frac{\alpha}{1+\alpha}\sin\left(\pi\frac{\Phi_{ex}-\Phi_\pi}{\Phi_0}\right)^2 \sin2\phi
	\right],\label{current}
\end{align}
where $\phi$ $\equiv$ $(\phi_a+\phi_b)/2$\cite{Romeo05}. 
Notably, the second term in Eq.(\ref{current}) including $\lq\lq\sin 2\phi$" is the 
origin of the half-integer SS.
By applying a voltage $\tilde{V}(t)$ = $V$ + $V_1\cos\Omega t$, $\phi$ = $\phi(0)$ + $a t$ + $b\sin\Omega t$ with $a$ = $2\pi V/\Phi_0$ and $b$ = $2\pi V_1/(\Phi_0\Omega)$, $i_B$ becomes,
\begin{align}
	i_B&\sim
	4\frac{\alpha}{1+\alpha}{\rm Im}\left[A e^{i\phi(0)}\sum_k J_k(b)e^{i(a+k\Omega) t}\right.\nonumber\\
	&\left.+\frac{\beta}{2}\frac{\alpha}{1+\alpha} B^2 e^{2i\phi(0)}
	\sum_{k'} J_{k'}(2b)e^{i(2a+k'\Omega) t}\right],\label{ib}
\end{align}
where $A$ $\equiv$ $\cos\left(\pi\frac{\Phi_{ex}+\Phi_\pi}{\Phi_0}\right)$, $B$ $\equiv$ $\sin\left(\pi\frac{\Phi_{ex}+\Phi_\pi}{\Phi_0}\right)$, and the $k$th order Bessel function $J_k(b)$. 
For the $\pi$-SQUID with $\Phi_{ex}$ = 0, the first term in Eq. (\ref{ib}) is zero because $\Phi_\pi/\Phi_0$ = 1/2. 
When $V$ satisfies $ V/(\Omega\Phi_0/2\pi)$ = $k'/2$, the SS with a half-integer $k'/2$ and an integer $q$ ($k'$ = 2$q$) appear with a DC-component, 
$2\beta[\alpha/(1+\alpha)]^2 J_{k'}(2b)$, for $\phi(0)$ = $\pi/4$. 
Meanwhile, in the conventional SQUID with $\Phi_{ex}$ = 0, the second term in Eq. (\ref{ib}) is zero because $\Phi_\pi/\Phi_0=0$. 
Only integer SS appear at voltages of integer multiples of  
$V/(\Omega\Phi_0/2\pi)=k$ with 
a DC-component, 
$4[\alpha/(1+\alpha)]J_k(b),$ for $\phi(0)$ = $\pi/2$. 

\begin{figure}[thb]
	\centering
	\includegraphics[width=0.5\textwidth]{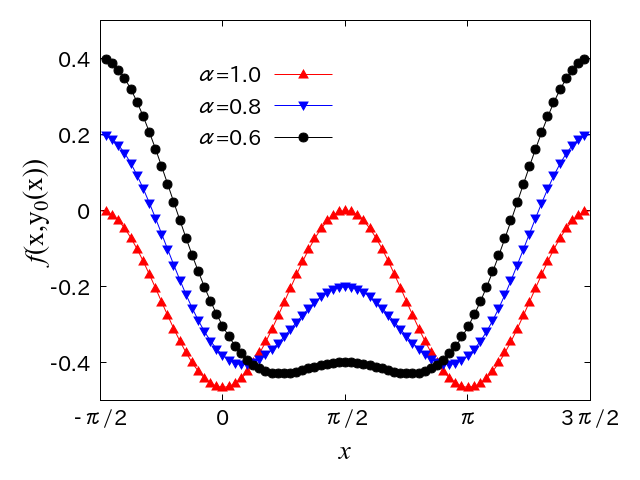}
	\caption{The potential energy for $\beta$ = 1.0 with $\alpha$ = 1.0 (red upper triangles), $\alpha$ = 0.8 (blue lower triangles), and $\alpha$ = 0.6 (black circles).}
	\label{qubit}
\end{figure}

The half-integer SS are explained by the onset of spontaneous 
current, leading to a flip-flop between two fluxoid states of the 
$\pi$-SQUID synchronized to the alternating field\cite{vanneste,frolov}. It suggests 
that the present system can become a qubit. 
Equation (\ref{current}) shows that the two potentials with $\phi$ and 2$\phi$ are convoluted similar to the $sds$-wave JJ\cite{ioffe99,shulz,lindstrom}. 
The half-integer SS and $\pi$-qubit are the two sides of the same 
coin.
The potential energy of the $\pi$-SQUID $U(\phi_a,\phi_b)$ with Eq. (\ref{4}) is given by, 
\begin{align}
	U(\phi_a,\phi_b)
		&=\frac{\Phi_0}{2\pi}\left[-J_a \cos \phi_a - J_b \cos(\phi_b)\right] +\frac{1}{2L}\Phi^2,\label{upot}\\
		&\equiv \left(J_a\frac{\Phi_0}{2\pi}\right)f(x,y),\\
	f(x,y)
		&= -\sin(x+y)+\alpha\sin(x-y)+\frac{2}{\beta}y^2,\label{pot}
\end{align}
where $x$ $\equiv$ $\phi$, and $y$ $\equiv$ $(\pi/\Phi_0) \Phi$. 
The ground state is obtained by minimizing $f(x,y)$ with respect to $y$ for a fixed $x$.
\begin{align}
	\left.\frac{\partial f(x,y)}{\partial y}\right|_{y=y_0}
		&= -\cos\left(x+y_0\right) -\alpha\cos\left(x-y_0\right) +\frac{4}{\beta}y_0=0,\label{const}
\end{align}
by which $y_0$ is determined as a function of $x$, i.e., $y_0$ = $y_0(x)$.
Equation (\ref{const}) means that $f(x,y)$ is minimized with respect to 
$\Phi_{ex}$, since we study the $\pi$-SQUID with two JJs shown in Fig. 
\ref{circuit}. To avoid $\Phi_{ex}$, the minimization with respect to 
$\Phi_{ex}$ can be substituted by another Josephson phase including one more JJ 
in the $\pi$-SQUID, as discussed in the previous 
studies\cite{ioffe99,yamashita06}.   
Numerically solving Eq. (\ref{const}), we find that $f(x,y_0(x))$ has the 
double minimum with respect to $x$ as shown in Fig. \ref{qubit}. 
The right minimum corresponds to the current circulating state, whereas the left one has no circulating current. 
Similar to the previous case\cite{yamashita05}, the barrier height is 
suppressed by decreasing $\alpha$, which coincides with the suppression of the 
half-integer SS. When the barrier height was zero, the stable state 
is $x$ = $\pi$/2, which may be realized by setting the phase-lock to $\phi_a$ = 
$\phi_b$ using Eqs. (\ref{4}) and (\ref{flux}). In this case, the spontaneous 
loop current $I_a-I_b$ = 0. Then, the SS appear at voltages with 
integer multiples. The two minima correspond to the clockwise and 
anticlockwise loop currents in addition to the $\pi$-shift because of the 
$\pi$-JJ, meaning that the spontaneous loop current is induced. Then, $\phi_a$ 
and $\phi_b$ are not synchronized as discussed in Refs. \cite{vanneste} 
and \cite{frolov}, because the loop current means the time-evolution of the 
phase difference in each junction. 

\begin{figure}[htbp]
	\centering
	\includegraphics[width=0.5\textwidth]{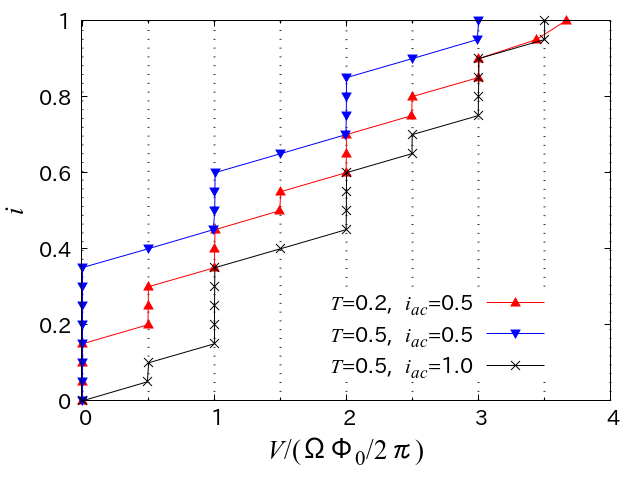}
	\caption{$T$-dependence of $I$$-$$V$ curve with $\Omega/\omega_0$ = 0.1, $\alpha$ = 1.0, and $\beta$ = 1.0. Each line is obtained for 
		$T$ = 0.2 and $i_{ac}$ = 0.5 (red upper triangles), 
		$T$ = 0.5 and $i_{ac}$ = 0.5 (blue lower triangles), and 
		$T$ = 0.5 and $i_{ac}$ = 1.0 (black crosses).}
	\label{metaliv}
\end{figure}

It is useful to compare the present model with the previous one, in which a metallic transport is assumed in the $\pi$-JJ\cite{yamashita05}.
The Josephson current of the metallic junction assigned to the junction-$b$ is given by,
\begin{equation}
	J_b \frac{\sin\phi_b}{\sqrt{1-T\sin^2(\phi_b/2))}},
\end{equation}
with transmittance $T$ (0 $\leq T\leq$ 1)\cite{kulik,bardeen,furusaki}.
This case also shows the half-integer SS as shown in Fig. \ref{metaliv}, 
although its magnitude becomes small when $T$ approaches 1, where the magnitude 
of the metallic junction becomes large and dominates the current. 
Even in such a metallic case, the half-integer SS can be observed by 
increasing the $i_{ac}$. 

\begin{figure}[htbp]
	\centering
	\includegraphics[width=0.5\textwidth]{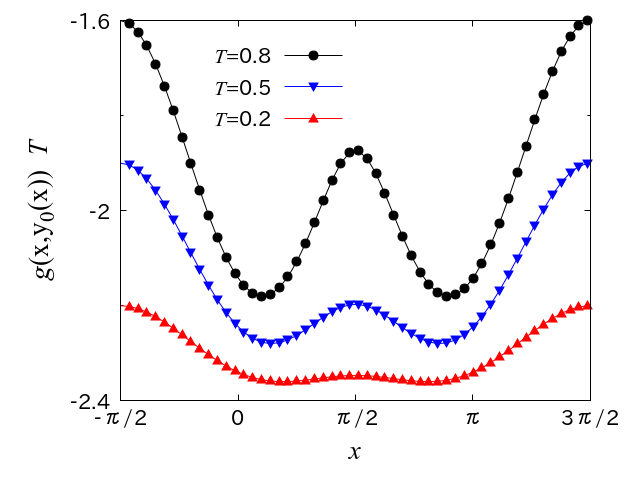}
	\caption{Potential energy for $\alpha$ = 0.6 and $\beta$ = 1.0 with $T$ = 0.8 (black circles), $T$ = 0.5 (blue lower triangles), and $T$ = 0.2 (red upper triangles).}
	\label{metalpotential}
\end{figure}

The potential energy corresponding to Eq. (\ref{pot}) is given by,
\begin{align}
	g(x,y)
	&= -\sin(x+y)+\frac{4\alpha}{T} \sqrt{1-\frac{T}{2}\left[1+\sin\left(x-y\right)\right]}+\frac{2}{\beta}y^2.\label{pot2}
\end{align}
The ground state is obtained by minimizing $g(x,y)$ with respect to $y$ for a fixed $x$.
\begin{align}
	\left.\frac{\partial g(x,y)}{\partial y}\right|_{y=y_0}
		&= -\cos\left(x+y_0\right) 
			-\frac{\alpha\cos\left(x-y_0\right)}{\sqrt{1+\frac{T}{2}\left[1+\sin\left(x-y_0\right)\right]}} +\frac{4}{\beta}y_0=0.\label{const2}
\end{align} 
Numerically solving Eq. (\ref{const2}), we find that $g(x,y_0(x))$ exhibits the 
double minimum with respect to $x$ as shown in Fig. \ref{metalpotential}. 
Because the magnitude of potential highly depends on $T$, $g(x,y_0(x))\times T$ is 
plotted instead of $g(x,y_0(x))$ for clarity.
The rather insulating case with $\alpha$ = 0.6, which corresponds to the black circles in Fig. \ref{qubit}, shows the shallow minimum, whereas the double minimum becomes clear by increasing $T$; this would contradict to the suppression of the half-integer 
SS in Fig. \ref{metaliv}, i.e., red upper triangles and blue lower 
triangles. However, it is caused by a large magnitude of potential in the metallic 
junction. In fact, the half-integer SS are revived by 
increasing $i_{ac}$. 

In this paper, we assumed that the Josephson critical current densities $j_i$ ($i=a,b$) defined by $j_i\equiv J_i/A_i$ with junction area $A_i$ are common between junctions-$a$ and $b$, i.e., $j_b/j_a=1$. Even in the case of $j_b/j_a \ne 1$, the half-integer SSs can remain for $0.7\lesssim j_b/j_a\lesssim 1.3$. We can consider another type of SQUID including two 0-JJs and one $\pi$-JJ, which is experimentally realized\cite{feofanov}. In this case, the $\pi$-JJ is used as a $\pi$-phase shifter. The condition is quite similar to the SQUID with an external flux studied by Vanneste et al. \cite{vanneste}. We can expect the half-integer SSs in such a geometry as well. 

So far, we have focused on the overdamped JJs to clarify the close relation between the half-integer SSs and the $\pi$-qubit. From a viewpoint of qubit application and/or operation, on the other hand, we need to include the capacitance in Eq. (\ref{rsj}) and should estimate the coherence time of qubit. In fact, Kato et al., discussed a long coherence time in a $\pi$-qubit \cite{kato}. Although the coherence time of qubit is a crucial factor, it goes beyond our aim in this paper. We will examine the coherence time of $\pi$-SQUID by considering a capacitance and thermal fluctuations in the near future.

In summary, the half-integer SS in $\pi$-SQUID comprising 0-and $\pi$-JJs have 
been studied using the RSJ model. 
We have shown that the $\pi$-SQUID can be a $\pi$-qubit with spontaneous loop currents, by which 
the half-integer SS are induced, meaning that the half-integer SS and the $\pi$-qubit are the two sides of the same 
coin. 
Making the 0- and $\pi$-JJs equivalent is a key for the half-integer Shapiro 
steps and realizing the $\pi$-qubit.

\begin{acknowledgements} 
We thank Wei Han and Yunyan Yao for their useful discussions. 
This work was supported by JSPS Grant Nos.~JP20K03810 and JP21H04987, and the inter-university cooperative research program (No.~202012-CNKXX-0008) of the Center of Neutron Science for Advanced Materials, Institute for Materials Research, Tohoku University. A part of the computations was performed on supercomputers at the Japan Atomic Energy Agency. 
S.M. is supported by JST CREST Grant (Nos. JPMJCR19J4, JPMJCR1874, and JPMJCR20C1)
and JSPS KAKENHI (nos. 17H02927 and 20H01865) from MEXT, Japan.
\end{acknowledgements}

\end{document}